\documentclass[a4paper,twoside,10pt]{article}

\input{jgrg19.sty}
\usepackage{bm}
\def\bm#1{{\boldsymbol{#1}}}

\begin{document}
%
\pagestyle{fancy}
\fancyhead{}
  \fancyhead[RO,LE]{\thepage}
  \fancyhead[LO]{H. Goto}                  
  \fancyhead[RE]{Lensing in LTB}    
\rfoot{}
\cfoot{}
\lfoot{}
\label{P6}    
\title{%
  Gravitational Lensing Effects in the LTB Model
}
%
\author{%
  Hajime Goto\footnote{Email address: gotohaji@post.kek.jp}$^{(a)}$
  and
  Hideo Kodama$^{(a),(b)}$
}
%
\address{%
  $^{(a)}$Department of Particle and Nuclear Physics, Graduate University for Advanced Studies,\\
       Tsukuba, Ibaraki 305-0801\\
  $^{(b)}$KEK, Tsukuba, Ibaraki 305-0801}
%
\abstract{
In this talk, we discuss how to estimate the gravitational lensing effect of a local void on the CMB polarization by using the LTB model.
}

\section{Introduction}
\label{sec:P6_intro}

Type Ia supernova (SNIa) observations imply an acceleration of the cosmic expansion if the universe is homogeneous and isotropic on scales larger than 200Mpc. If we abandon this assumption called the Cosmological Principle, then other explanations become possible. The most interesting model of such a nature is the local void model, which was first proposed by Kenji Tomita in 2000 \citep{2000}.  This model assumes that we are around the center of a low density spherically symmetric void and the spacetime is well described by the Lema\^{i}tre-Tolman-Bondi (LTB) model. In this model, the cosmic expansion rate decreases outward at each constant time slice, which produces an apparent acceleration of the universe when observed along the past light cone. Although this model violates the Cosmological Principle and requires the accidental situation concerning our location in the universe, it does not require any dark energy or a modification of gravity theory. Further, as far as the redshift-luminosity distance relation obtained by the SNIa observations is concerned, this model can reproduce the observational results with any accuracy because it contains at least one arbitrary function of the radius. Actually, it has passed all observational tests so far. Therefore, it is of crucial importance to find observational tests that enable us to discriminate this void model from the FLRW-based models, in order to establish the necessity of dark energy or a modification of gravity.

One possible such test is to observe  effects of the inhomogeneity on CMB temperature and polarization. For example, gravitational lensing is expected to generate B-mode and the $\langle$EB$\rangle$ correlation for an off-center observer in the local void model. In this talk, we explain how to calculate such gravitational lensing effects on CMB in the LTB model.

\section{CMB Polarization}
\label{sec:P6_pol}

\subsection{How to Represent Polarizations}
\label{sec:P6_rep}

First of all, we explain the standard method to represent the polarization of radiations. Let us consider an quasi-monochromatic plane electromagnetic wave propagating toward an observer, and take an orthonormal $xy$-basis that is orthogonal to the wave propagation direction. Then, the electric field of the wave is represented as $\bm{E} = E_x \bm{e}_x + E_y \bm{e}_y$, with $E_x = a_x \sin (\omega t - \epsilon_x)$ and $E_y = a_y \sin (\omega t - \epsilon_y)$.

In this setup, one can define parameters that represent polarization as follows: $I := \langle a_y^2 \rangle + \langle a_x^2 \rangle$, $Q := \langle a_y^2 \rangle - \langle a_x^2 \rangle$, $U := \langle 2 a_y a_x \cos (\epsilon_y - \epsilon_x) \rangle$, and $V := \langle 2 a_y a_x \sin (\epsilon_y - \epsilon_x) \rangle$.
These are called the Stokes parameters.
Physically, $I$ represents intensity (temperature), $Q$ and $U$ represent linear polarization, and $V$ represents circular polarization.
We ignore $V$ because circular polarization is never generated by Thomson scattering in the early universe.

If the orthonormal basis is rotated in the wave plane, $Q$ and $U$ are linearly transformed. Since this is inconvenient, we will introduce new quantities below that are independent of the choice of the orthonormal basis.

\subsection{Polarization Distribution Patterns}
\label{sec:P6_pat}

Up to this point, we have considered a wave propagating only in one direction, whose polarization can be described by the Stokes parameters $(I, Q, U)$. In real observations, this set of parameters is measured for photons of each direction, and the result is represented by three functions on the sky, $I(\boldsymbol{\theta}_\mathrm{obs})$, $Q(\boldsymbol{\theta}_\mathrm{obs})$ and $U(\boldsymbol{\theta}_\mathrm{obs})$, where $\boldsymbol{\theta}_\mathrm{obs}$ represents the position on the sky. 

In the present article, for simplicity, let us work in the flat-sky approximation. This approximation is valid when we consider only a small part of the whole sky. 
Let us define a tensor from $Q$ and $U$ as
\begin{equation} \label{equ:poltensor} P_{ab} (\bm{x}) = \frac{1}{2}
\left( \begin{array}{*{2}{c}}
Q(\bm{x}) & U(\bm{x}) \\
U(\bm{x}) & -Q(\bm{x})
\end{array} \right) \, ,
\end{equation}
where the subscripts of $P$ run over $x$ and $y$ ($x \equiv 1$, $y \equiv 2$).
This tensor is called the polarization tensor.

The polarization tensor field can be used to define the two functions on the sky, $E(\bm{x})$ and $B(\bm{x})$, which are independent of the choice of the orthonormal basis, by $\nabla^2 E (\bm{x}) = \partial_{a} \partial_{b} P_{ab} (\bm{x})$, $\nabla^2 B (\bm{x}) = \epsilon_{ac} \partial_{b} \partial_{c} P_{ab} (\bm{x})$. 
Because these represent the `gradient' and `curl' (or `rotation') components of the linear polarization distribution, respectively, they are called the E-modes and the B-modes, respectively. 

The power spectra and correlation functions are defined in terms of their Fourier transformations $\tilde{E} (\bm{\ell})$ and $\tilde{B} (\bm{\ell})$ as 
\begin{equation} \label{equ:powerspectrum} \langle \tilde{X}_1 (\bm{\ell}) \tilde{X}_2 (\bm{\ell}') \rangle
= (2 \pi)^2 \delta(\bm{\ell} + \bm{\ell}') C^{\mathrm{X}_1 \mathrm{X}_2}_\ell \, ,
\end{equation}
with $\tilde{X}_1,\ \tilde{X}_2 \in \{ \tilde{\Theta},\ \tilde{E},\ \tilde{B} \}$ and $\mathrm{X}_1,\ \mathrm{X}_2 \in \{ \mathrm{T},\ \mathrm{E},\ \mathrm{B} \}$, where $\Theta$ represents the intensity (temperature) fluctuation around the sky average (`2.7K').
If physics and the ensemble for averaging are invariant under a parity inversion, it turns out that $C^{\mathrm{TB}}_\ell =C^{\mathrm{EB}}_\ell = 0$.


In the real, spherical-sky case, a similar argument holds.
For details, the reader is referred to Ref.~\citep{2004astro.ph..3392C}.

\section{Gravitational Lensing Effects}
\label{sec:P6_gl}

Inhomogeneous gravitational fields produce two effects on photon propagation. The first is a bending of its trajectory, and the second is the change of the photon energy in addition to the standard redshift by cosmic expansion. The latter is the so-called Sachs-Wolfe effect, which we do not consider in this article. Intuitively speaking, as far as CMB measurements by a fixed observer are concerned, the former---called `shear field effect'---can be further divided into two parts: ({\em{i}}) the change of the photon direction in the sky and ({\em{ii}}) the displacement of the intersection sphere of the past light cone and the last scattering surface in the direction perpendicular to this sphere. In order to give a definite meaning to this distinction, we need to introduce some reference FLRW model to define `unperturbed' photon trajectories and past light cones. However, this procedure introduces the gauge freedom corresponding to the mapping between the real universe and the reference model, and thus make that distinction obscure.  In fact, for the FLRW model with small perturbations, the displacement of the last scattering sphere can be set to be zero by an appropriate gauge choice, and in this gauge, the shear field effect can be represented only in terms of ({\em{i}}), namely, the `gravitational lensing effect'. In the local void model, it is not so certain whether the same argument holds when the non-linearity of inhomogeneities is large. In the present article, we simply assume that the shift of the last scattering point in the direction normal to the last scattering sphere can be set to zero by a gauge choice. 

\subsection{General Formula}

\begin{figure}[t]
\centering
\includegraphics[keepaspectratio=true,width=10cm, trim=100 50 100 100]{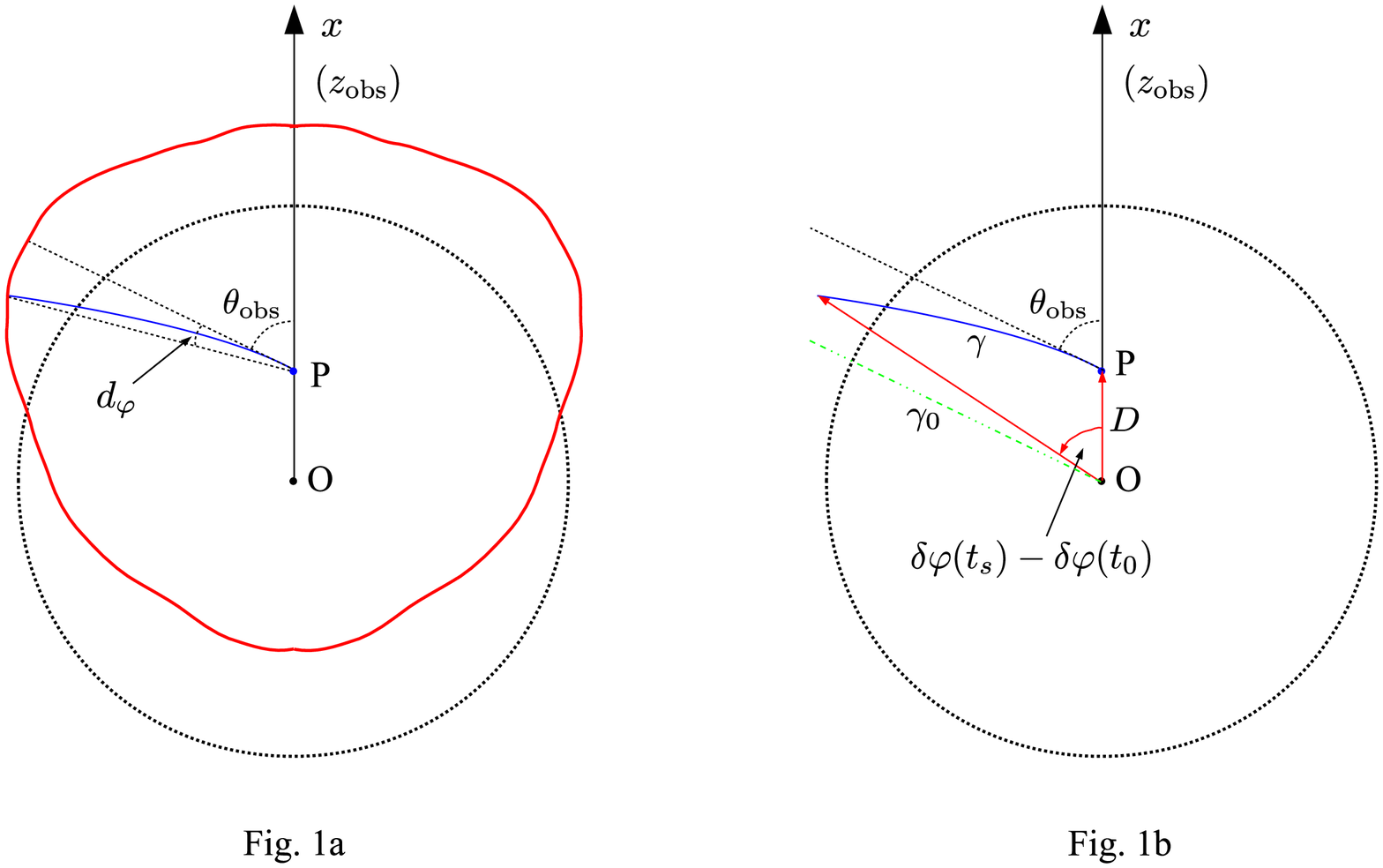}
\caption{
(a) Photon propagation in the LTB model. The black circle (dashed line) represents the last scattering sphere for the observer at the center O, and the red curve (solid line) represents that for an off-center observer P. The blue curve (arriving at P) is the photon trajectory. (b) The photon trajectory $\gamma$ in the $(r,\varphi)$ coordinates. Each trajectory is contained in the unique two-plane passing through O, P and the last scattering point. The green line (dashed-dotted line) represents the reference radial null geodesic $\gamma_0$.
}
\label{fig:P6_drawing}
\end{figure}

Under this assumption, the gravitational lensing effect on the CMB anisotropy can be simply determined by the two-dimensional shift vector $\bm{d}$ on the sky representing the difference between the observed direction of a photon and its initial direction on the last scattering sphere (Fig.~\ref{fig:P6_drawing}a). The same formula holds for the LTB model as that for the FLRW model~\citep{1996ApJ...463....1S}, \citep{1998PhRvD..58b3003Z}, \citep{2000PhRvD..62d3007H}, and \citep{2006PhR...429....1L}.

In the flat-sky approximation, let us represent the CMB temperature and polarization that would be observed in the reference exact FLRW model without gravitational lensing as $\tilde{\Theta}$, $\tilde{Q}$ and $\tilde{U}$ ($\tilde{E}$ and $\tilde{B}$). Then, the corresponding quantities observed with gravitational lensing are given by
\begin{eqnarray}
\Theta(\bm{\theta}_{\mathrm{obs}}) 
&=& \tilde{\Theta}(\bm{\theta}_{\mathrm{obs}} + \bm{d}) 
   = \int \frac{d^2 \bm{\ell}}{(2\pi)^2} e^{i\bm{\ell}\cdot(\bm{\theta}_{\mathrm{obs}}
     + \bm{d})} \tilde{\Theta}(\bm{\ell}) \ , 
\label{eqn:P6_T} \\
Q(\bm{\theta}_{\mathrm{obs}}) 
&=& \tilde{Q}(\bm{\theta}_{\mathrm{obs}}+ \bm{d}) 
   = \int \frac{d^2 \bm{\ell}}{(2\pi)^2} e^{i\bm{\ell}\cdot(\bm{\theta}_{\mathrm{obs}}
     + \bm{d})}\{\tilde{E}(\bm{\ell})\cos2\phi_{\bm{\ell}} 
     - \tilde{B}(\bm{\ell})\sin2\phi_{\bm{\ell}}\} \ , 
\label{eqn:P6_Q}\\
U(\bm{\theta}_{\mathrm{obs}}) 
&=& \tilde{U}(\bm{\theta}_{\mathrm{obs}}+ \bm{d}) 
  = \int \frac{d^2 \bm{\ell}}{(2\pi)^2} e^{i\bm{\ell}\cdot(\bm{\theta}_{\mathrm{obs}}
     + \bm{d})} \{ \tilde{E}(\bm{\ell})\sin2 \phi_{\bm{\ell}} 
      + \tilde{B}(\bm{\ell}) \cos 2\phi_{\bm{\ell}} \} \ ,
\label{eqn:P6_U}
\end{eqnarray}
where $\phi_{\boldsymbol{\ell}}$ is the angle between the vector $\bm{\ell}$ and the $x$-direction on the sky. From this, the Fourier transformation of the differences are
\begin{eqnarray}
\delta\Theta(\bm{\ell}) &=& \int \frac{d^2 \bm{\ell}'}{(2\pi)^2} \tilde{\Theta}(\bm{\ell}') W(\bm{\ell}', \bm{L}) \ , 
\label{eqn:P6_deltaT} \\
\delta E(\bm{\ell}) &=& \int \frac{d^2 \bm{\ell}'}{(2\pi)^2} \{ \tilde{E} (\bm{\ell}') \cos 2 \phi_{\bm{\ell}'\bm{\ell}} - \tilde{B}(\bm{\ell}') \sin 2 \phi_{\bm{\ell}'\bm{\ell}} \} W(\bm{\ell}', \bm{L}) \ , 
\label{eqn:P6_deltaQ} \\
\delta B(\bm{\ell}) &=& \int \frac{d^2 \bm{\ell}'}{(2\pi)^2} \{ \tilde{E}(\bm{\ell}') \sin2 \phi_{\bm{\ell}'\bm{\ell}} + \tilde{B} (\bm{\ell}') \cos 2 \phi_{\bm{\ell}'\bm{\ell}}  \} W(\bm{\ell}', \bm{L}) \ , 
\label{eqn:P6_deltaU} 
\end{eqnarray}
where $W(\bm{\ell}, \bm{L}) = - \bm{\ell}\cdot \bm{d}(\bm{L})$, $\bm{L}= \bm{\ell} - \bm{\ell}'$, $\phi_{\bm{\ell}'\bm{\ell}} := \phi_{\bm{\ell}'} - \phi_{\bm{\ell}}$, and $\bm{d}(\bm{L})$ is the Fourier transformation of $\bm{d}(\bm{\theta}_\mathrm{obs})$.


\subsection{Null Geodesics in the LTB Model and the Shift Vector $\pmb{d}$}
\label{sec:P6_ge}

Thus, the investigation of the gravitational lensing effect of a local void on CMB is reduced to determine $\bm{d}$ as a function of the photon direction. For that, we have to solve the null geodesic equation in the LTB model, whose metric can be written $ds^2 = -dt^2 + S^2 dr^2 + R^2(d\theta^2 + \sin^2 \theta d\varphi^2)$. Here $R$ is a function of $t$ and $r$, and $S$ is written in terms of $R$ and the curvature function $k(r)$ as $S=R'/(1-k(r)r^2)^{1/2}$.

In terms of the photon 4-momentum $p^\mu=dx^\mu/d\lambda$ with affine parameter $\lambda$, the geodesic equation can be written as ${d p^\mu}/{d \lambda} = - \Gamma^\mu_{\nu\rho} p^\nu p^\rho$. Because of the spherical symmetry, this set of equations can be reduced to the coupled ODEs for $\omega$, $\mu$ and $p_\perp$ defined by $p^t = \omega$, $p^r = \mu \omega /{S}$, and $p_\perp^2 = \omega^2 (1-\mu^2)$, where $p_\perp := R \{( p^{\theta})^2 + (p^{\varphi})^2 \sin^2 \theta \}^{1/2}$. Without loss of generality, we can assume that the photon propagate on the 2-plane with $\theta=\pi/2$, and therefore $p^\theta=0$ (Fig.~\ref{fig:P6_drawing}b). Then, the geodesic equations are reduced to the set of four ODEs for $\omega(t)$, $r(t)$, $\varphi(t)$, and $\mu(t)$.


It is impossible to express the general solution to this set of equations in terms of known functions. However, if we restrict the consideration to null geodesics passing through a point P close to the symmetry center O, we can find explicit expression for the solution in terms of integrals of known functions. Such a geodesic $\gamma$ stays close to some radial null geodesic $\gamma_0$. Hence, we can solve the geodesic equation perturbatively with respect to the deviation of the two geodesics $\gamma$ and $\gamma_0$. One subtle point in this perturbative approach is that $\delta \mu$ turns out to show bad behavior near the observer. This problem can be avoided by introducing the variables $b$ \& $c$ defined as $b := r \sqrt{1 - \mu^2}$ and $c := r \mu$ instead of $r$ and $\mu$.

The final result reads
%
\begin{eqnarray} \label{eqn:P6_bcdeltac} \delta c (t) &=& \delta c (t_0) \exp \int^t_{t_0} dt_1 \left(\frac{S'}{S^2}\right)_{t_1} \ , \\
 \delta \omega (t) &=& \left[ \delta \omega (t_0) - \delta c (t_0) \int^t_{t_0} dt_1 \left\{ \omega \left( - \frac{\dot{S}'}{S} + \frac{\dot{S}S'}{S^2} \right) \right \}_{t_1} \exp \int^{t_1}_{t_0} dt_2 \left( \frac{S'}{S^2} + \frac{\dot{S}}{S} \right)_{t_2} \right] \nonumber \\
& & \cdot \exp \left(-\int^t_{t_0} dt_1 \left(\frac{\dot{S}}{S} \right)_{t_1} \right) \ ,
\label{eqn:P6_bcdeltaomega} \\
\label{eqn:P6_bcdeltab} \delta b (t) &=& \delta b (t_0) \exp \int^t_{t_0} dt_1 \left( \frac{R'}{SR} - \frac{\dot{R}}{R} + \frac{\dot{S}}{S} + \frac{1}{cS} \right)_{t_1} \ , \\
\label{eqn:P6_bcdeltavarphi} \delta \varphi (t) &=& \delta \varphi (t_0) \pm \delta b (t_0) \int^t_{t_0} dt_1 \left( \frac{1}{|c|R} \right)_{t_1} \exp \int^{t_1}_{t_0} dt_2 \left( \frac{R'}{SR} - \frac{\dot{R}}{R} + \frac{\dot{S}}{S} + \frac{1}{cS} \right)_{t_2} \ . \end{eqnarray}


In Fig.~\ref{fig:P6_drawing}b, we can take the null geodesic passing through O in the direction $\theta_{\rm obs}$ as the reference geodesic $\gamma_0$. Then, it is easy to see that the values of $\delta b$ and $\delta c$ at present $t=t_0$ can be expressed as
\begin{equation}
\delta b(t_0)= D \sqrt{1-\mu_0^2} \ , \quad
\delta c(t_0)= D\mu_0 \ , 
\end{equation}
where $D$ is the distance of the observer P from O and $\mu_0=-\cos(\theta_{\rm obs})$. Then, 
\eqref{eqn:P6_bcdeltavarphi} determines the shift vector $\boldsymbol{d} = ( d_\theta, d_\varphi )$, $d_\theta = 0$, $d_\varphi = \delta\varphi(t_s) - \delta\varphi(t_0)$, where $t_s$ is the last scattering time.

\section{Summary and Future Work}
\label{sec:P6_sum}
In this paper, we developed a formulation to calculate the gravitational lensing effects on the CMB temperature and polarization for an off-center observer in a spherically symmetric void described by the LTB model. Explicit estimations of these effects are under investigation \citep{0000}.

\section{Acknowledgements}
\label{sec:P6_acknowledge}

The authors would like to thank all participants of the workshop
$\Lambda$-LTB Cosmology (LLTB2009) held at KEK from 20 to 23  October
2009 for useful discussions.  This work is supported by the project
Shinryoiki of the SOKENDAI Hayama Center for Advanced Studies and the
MEXT Grant-in-Aid for Scientific Research on Innovative Areas (No.
21111006).



\end{document}